# Electrical transport properties of RNiO$_3$ (R= Pr, Nd, Sm) Epitaxial Thin Films

A. Venimadhav, I. Chaitanyalakshmi and M.S. Hegde

Solid State and Structural Chemistry Unit, Indian institute of Science, Bangalore-560012.

Electrical transport properties of RNiO$_3$ (R= Pr, Nd, Sm) thin films grown by pulse laser deposition have been studied. RNiO$_3$ films grow in (100) direction on LaAlO$_3$ (100) substrate. Unlike in polycrystalline solid, PrNiO$_3$ film showed metallic behavior. The first order metal to insulator transition observed in polycrystalline solids is suppressed in RNiO$_3$ films. The effect of lattice strain in the films influensing the transport properties has been studied by varying the thickness of PrNiO$_3$ film on LaAlO$_3$ and also by growing them on SrTiO$_3$ and sapphire substrates. Deviation in the transport properties is explained due to the strain induced growth of the films. Further, we show that the transport property of LaNiO$_3$ film is also influenced by a similar strain effect.

mshegde@sscu.iisc.ernet.in
madhav@sscu.iisc.ernet.in

Metallic oxides crystallizing in perovskite related structures have got potential application in various fields such as microelectronics, electrodes for thin film devices and heterostructure devices. Amongst the perovskite metallic oxides, $LaNiO_3$ exhibits high electrical conductivity without doping.[1] Epitaxial thin film of $LaNiO_3$ has been made by pulsed laser deposition (PLD) and its metallic behavior is well-studied.[2] The other members of the series, $RNiO_3$ (R= Pr, Nd and Sm) are found to crystallize in the $GdFeO_3$- type orthorhombic distorted perovskite structure.[3] These compounds exhibit first order metal to insulator transition on cooling. As the rare earth size is increased the compounds become more conducting and the metal to insulator transition (MI) temperature decreases from $SmNiO_3$ (400K) to $PrNiO_3$ (130) and becomes fully metallic in $LaNiO_3$.

Rare earth nickelates are very important in order to understand the physics of oxide perovskites as they are in the boundary between low-$\Delta$ (charge transfer energy) metals and charge transfer insulator phases.[4] The specialty of the nickelates is that the MI transition is not accompanied by change in lattice symmetry and magnetic transition. The application of hydrostatic pressure suppresses the MI transition in $RNiO_3$.[5,6] The application of external pressure brings the metallic behavior in $PrNiO_3$ down to low temperatures.

Among $RNiO_3$, only $LaNiO_3$ could be prepared at atmospheric pressure of oxygen. Other members are prepared under high oxygen pressure to stabilize Ni in 3+ oxidation state. However PLD has been proven to be a convenient method to grow epitaxial films stabilizing the transition metals in higher oxidation state. For example the metallic $LaCuO_{3-\delta}$ epitaxial films were made by PLD under normal deposition conditions.[7] Thin films NdNiO$_3$ has showed metal to insulator property.[8,9] Here we report the transport property of the thin films of RNiO$_3$ (R = Pr, Nd, Sm) by PLD. We have found that PrNiO$_3$ shows metallic behavior. The first order MI transition is suppressed in the RNiO$_3$ films. The deviation of transport properties from the bulk solids to thin films has been explained on the basis of strain induced growth of the film.

Pr$_6$O$_{11}$ and NiO were mixed in atomic ratios of 1:1 for Pr:Ni and the pellet sintered at 1100 $^0$C was used as the target. The pellet predominantly consisted of Pr$_2$NiO$_4$ and NiO. Similarly R$_2$O$_3$ (R = Nd and Sm) and NiO were mixed in equal R:Ni atomic ratio and pellets were made at 1100 $^0$C. Thin films of RNiO$_3$ were prepared at a substrate temperature of 750 $^0$C in a flowing oxygen atmosphere of 330 mTorr. Thickness of the films were approximately 300 nm. Epitaxial films of LaNiO$_3$ (LNO) were also freshly made on LAO (100), SrTiO$_3$ (100) (STO) and sapphire for comparison. The out of plane lattice parameters were obtained by JEOL JDX8P x-ray diffractometer. Conductivity measurements were performed by four-probe method. The composition analysis by EDAX showed a 1:1 ratio of rare earth to nickel within the experimental error of 3%.

The XRD patterns showed highly oriented growth of the films. From the X-ray θ-2θ scans, the out of plane parameters (a$_0$) of PrNiO$_3$ (PNO), NdNiO$_3$ (NNO) and SmNiO$_3$ (SNO) films grown on LAO (100) have been calculated to be 3.83 Å, 3.81 Å and 3.80 Å respectively. Lattice parameters of RNiO$_3$ solids and out of plane parameter of the thin films are summarized in Table. 1. The observed a$_0$ values match closely with the 'a/√2' of their polycrystalline solids. Hence the growth of the films on LAO (100) substrate is along (100) direction. PNO film grown on STO (100) substrate also gave a$_0$ value of 3.83

Å close to the bulk 'a/√2' indicating a similar kind of (100) growth. The LNO film on LAO and STO also showed a (100) growth.

Resistivity Vs temperature behavior of PNO, NNO and SNO on LAO substrate is given in Fig .1. Unlike in polycrystalline solid, PNO film showed metallic behavior. NNO showed semimetallic and SMO exhibited semiconducting behavior. ρ Vs T of LNO film on LAO giving metallic behavior is also shown in the figure for comparison. From Fig. 1 it can be noted that as going from La to Sm the room temperature resistivity value increases from 70 μΩ cm (LNO), 3 mΩ cm (PNO), 5.2 mΩ cm NNO) to 11 mΩ cm (SNO). The MI transition expected for PNO, NNO and SNO in the polycrystalline solids is suppressed in the thin films. To understand the metallic behavior in PNO a more detailed study was undertaken.

PrNiO$_3$ film was grown on LAO ( $a_0$ = 3.788), STO ($a_0$ = 3.905) and sapphire ($a_0$ = 3.64) substrates to understand the effect of the lattice mismatch on the transport property. The ρ Vs T plots of PNO films on these substrates are shown in Fig 2. While PNO film is metallic on LAO, it showed a semimetallic behavior on STO and semiconducting behavior on sapphire. The room temperature resistivity on different substrates follow LAO<STO<sapphire. The resistivity values at 300K on these substrates were 3 mΩ cms, 5.4 mΩ cms and 11 mΩ cms respectively. Lattice mismatch of PNO ($a_0$=a/√2) is least on LAO (–1.13%) compared to STO (1.76%) and sapphire (-5.27%). Thus lower the lattice mismatch, higher the conductivity. This observation on PNO films led us to examine carefully the electrical transport properties of LNO on different substrates. ρ Vs T plots of LNO films are shown in the inset of Fig 2. LNO has least lattice mismatch on LAO (-0.71%) compared to SrTiO$_3$ (2.1%) and sapphire (-4.82%). Hence LNO is expected to be more metallic on LAO than on other two substrates. Indeed we have observed that, though resistivity behavior was metallic on both LAO and STO, the value of resistivity was found to be smaller on LAO (65 µΩcm) than on STO (170 µΩcm at 300K). It is noteworthy that the large lattice mismatch on sapphire induces metal-semiconductor behavior in $LaNiO_3$ thin film.

To investigate the effect of strain on the transport property of PNO film, the thickness dependent resistivity measurements were performed. Fig. 3 shows the resistivity Vs temperature behavior of PNO films on LAO with thicknesses of 75 nm, 300nm and 600nm. Film with smaller thickness showed a metallic behavior with a smaller resistivity value of 74 µΩcm and the 300 nm film still showed metallic behavior with a larger resistivity value of 3 mΩ cm. The 600 nm film showed a semiconducting behavior with 11 mΩ cm resistivity at room temperature showing a trend of reaching bulk transport property at higher thickness.

The interesting observation in the above results is the metallic behavior of the $PrNiO_3$ film and the suppression of the MI transition in RNO films. The transport properties of polycrystalline solids have not been reproduced in thin films. The strain produced by the lattice mismatch could be the reason for the above results. Several studies on the strain-property relations have been reported on perovskite oxides, particularly on rare earth manganites.[10-12] The thickness dependent strain effects have also been studied in manganites.[13-15] It is observed that with the increase of film thickness the strain relaxes and the transport properties tend to produce bulk solid behavior.

The metallic behavior in rare earth nickelates is believed to be due to charge transfer interaction between $Ni^{3+}$- O (2p). The crucial structural parameter that influence the electrical and magnetic properties in nicklates is Ni-O-Ni bond angle. The orthorhombic distortion in RNO (R = Pr, Nd, Sm) reduces the bond angle from that of the ideal cubic perovskite structure (i.e., $180^0$). The application of pressure reduces the volume of the cell such that the Ni-O-Ni bond angle stretches tending to $180^0$.[6] The metallic behavior in PNO film can be understood as follows. The growth of the film on LAO shows the out of plane parameter of 3.83 Å, which matches closely with 'a/√2' of bulk value. The other two reduced lattice parameters (3.81 Å, 3.80 Å) have to be accommodated on the 3.788 Å cubic lattice of LAO (100) plane. Hence a compressive strain in the b-c plane can shrink the volume of the cell as analogous to the hydrostatic pressure effect and drives to metallic behavior. This strain is large for smaller thickness and relaxes as the thickness of the film increases tending to the bulk transport behavior.

The metallic behavior of LNO was observed on both LAO and STO. This is true as the bond angle in LNO ($162.8^0$) is large enough to take care of the small strain variation on the two substrates, nevertheless the resistivity shows the clear distinction over the two substrates indicating LAO as a better substrate because of its smaller lattice mismatch. The metallic behavior not observed in NNO ($156.1^0$) and SNO ($152.6^0$) can be understood, as the bond angle is small compared to LNO and PNO ($158.7^0$).[5] Though metallic behavior was not observed in NNO and SMO, the strain suppresses the MI transition in these films. To gain a complete understanding of lattice strain effects on the transport properties, accurate in plane structural parameters are essential.

In conclusion, electrical transport properties of $RNiO_3$ films have been studied. PNO showed metallic behavior on LAO, semimetallic on STO and semiconducting on sapphire. The metal to insulator transition is suppressed in NNO and SMO films.

Resistivity values of metallic oxide films of LNO and PNO can be tailored by choosing substrates with appropriate lattice mismatch. The modulation of transport properties by the strain effect can be of technological importance in perovskite oxide heterostructures.

We greatly acknowledge the Department of Science and Technology and the Government of India for the financial support.

Table.1. Lattice parameters of $RNiO_3$ solids[3] and $a_0$ value of the films.

| Compound | a | a/√2 | b | b/√2 | c/2 | $a_0$ of the films |
|---|---|---|---|---|---|---|
| $PrNiO_3$ | 5.415 | 3.831 | 5.377 | 3.802 | 3.813 | 3.83 |
| $NdNiO_3$ | 5.3888 | 3.810 | 5.384 | 3.807 | 3.805 | 3.81 |
| $SmNiO_3$ | 5.366 | 3.794 | 5.452 | 3.855 | 3.790 | 3.80 |

Figure. 1 Resistivity Vs Temperature of $RNiO_3$ films.

Figure. 2 Resistivity Vs Temperature behavior of $PrNiO_3$ film on different substrates. Inset shows the resistivity Vs temperature behavior of $LaNiO_3$ film on different substrates.

Figure. 3 Resistivity Vs Temperature behavior of $PrNiO_3$ film with different thickness on $LaAlO_3$ substrate.

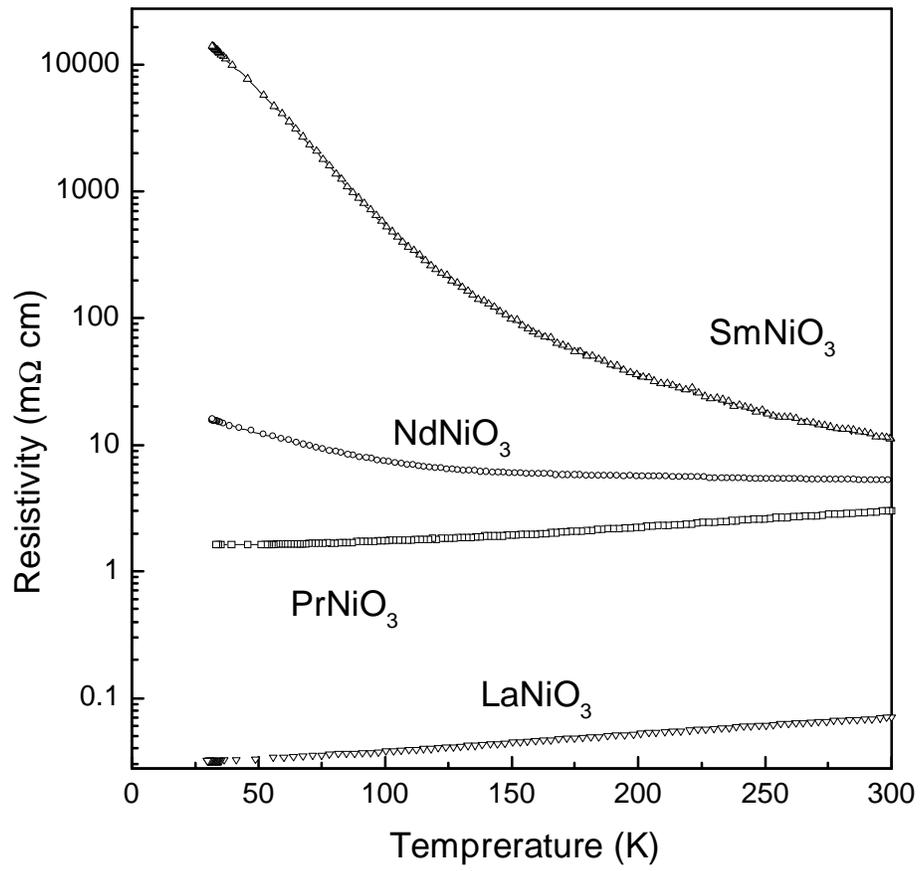

Fig. 1
A. Venimadhav et al.,

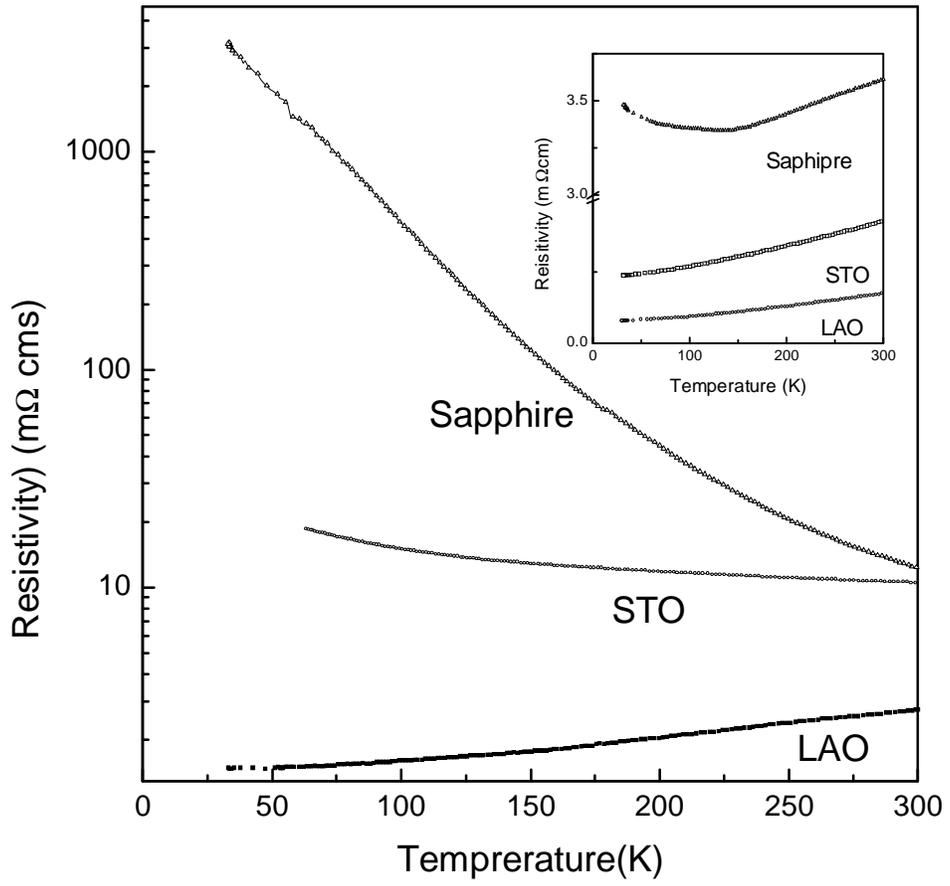

Fig. 2
Venimadhav et al

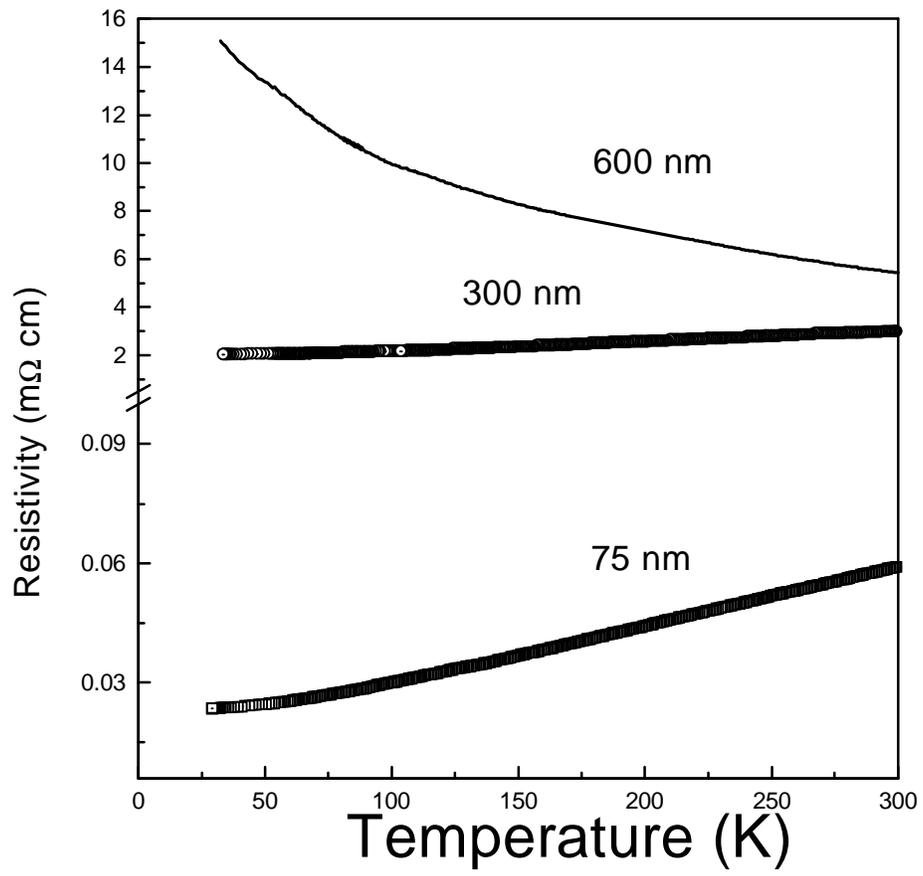

Fig. 3
Venimadhav et al